\documentclass[12pt,fleqn]{article}
\usepackage[DIV12]{typearea}
\usepackage{amsmath,amsfonts,amssymb,amscd}
\usepackage{graphicx}
\usepackage{cite}
\usepackage{mathrsfs}
\usepackage{bbm}
\usepackage{units}
\usepackage[small,loose]{subfigure}  
\usepackage{xcolor}
\usepackage{stmaryrd}
\usepackage{pifont}
\usepackage{hyperref}
\usepackage{cancel}

\newcommand{\Eqref}[1]{equation~\eqref{#1}}

\newcommand{\Tabref}[1]{table~\ref{#1}}

\newcommand{\Appref}[1]{appendix~\ref{#1}}

\newcommand{\SemiDirect}[0]{\ensuremath{\rtimes}}

\newcommand{\eVdist}{\kern-0.06em}

\DeclareMathOperator{\tr}{tr}

\DeclareMathOperator{\PL}{\ensuremath{\textit{P}_{\mathrm{L}}}}

\newcommand{\D}{\mathrm{d}}
\newcommand{\I}{\mathrm{i}}

\newcommand{\SO}[1]{\ensuremath{\mathrm{SO}(#1)}}
\newcommand{\SU}[1]{\ensuremath{\mathrm{SU}(#1)}}

\newcommand{\U}[1]{\ensuremath{\mathrm{U}(#1)}}
\newcommand{\Z}[1]{\ensuremath{\mathbbm{Z}_{#1}}} 

\newcommand{\rep}[2][]{\ensuremath{\boldsymbol{#2}#1}}

\allowdisplaybreaks[1]

\numberwithin{equation}{section}
\numberwithin{table}{section}

\definecolor{darkgreen}{HTML}{109930}

\def\mytitle{Anomaly--safe discrete groups}
\title{\mytitle}

\begin{document}

\begin{titlepage}

\begin{flushright}
 UCI-TR-2015-03\\
 TUM-HEP 991/15\\
 LMU-ASC 20/15\\
 FLAVOR-EU 96\\
 NSF-KITP-15-041
\end{flushright}

\vspace*{1.0cm}

\renewcommand*{\thefootnote}{\fnsymbol{footnote}}
\begin{center}
{\Large\textbf{\mytitle}}
\renewcommand*{\thefootnote}{\arabic{footnote}}

\vspace{1.0cm}

\textbf{Mu--Chun Chen\footnote[1]{Email: \texttt{muchunc@uci.edu}}{}$^a$,
Maximilian Fallbacher\footnote[2]{Email: \texttt{m.fallbacher@tum.de}}{}$^b$,
Michael Ratz\footnote[3]{Email: \texttt{michael.ratz@tum.de}}{}$^b$, \\
Andreas Trautner\footnote[4]{Email: \texttt{andreas.trautner@tum.de}}{}$^{b,c}$ 
and 
Patrick K.S.\ Vaudrevange\footnote[5]{Email: \texttt{patrick.vaudrevange@tum.de}}{}$^{c, d, e}$
}
\\[5mm]
\textit{$^a$\small
~Department of Physics and Astronomy, University of California,\\
Irvine, California 92697--4575, USA
}
\\[2mm]
\textit{$^b$\small
~Physik--Department T30, Technische Universit\"at M\"unchen, \\
James--Franck--Stra\ss e~1, 85748 Garching, Germany
}
\\[2mm]
\textit{$^c$\small
~Excellence Cluster Universe, \\
Boltzmannstra\ss e~2, 85748 Garching, Germany
}
\\[2mm]
\textit{$^d$\small
~TUM Institute for Advanced Study, \\
Lichtenbergstra\ss e~2a, 85748 Garching, Germany
}
\\[2mm]
\textit{$^e$\small
~Arnold Sommerfeld Center for Theoretical Physics,
Ludwig--Maximilians--Universit\"at M\"unchen, Theresienstra\ss e~37, 80333 M\"unchen, Germany
}
\end{center}

\vspace{5mm}

\begin{abstract}
We show that there is a class of finite groups, the so--called perfect groups, 
which cannot exhibit anomalies. This implies that all non--Abelian finite 
simple groups are anomaly--free. On the other hand, non--perfect groups 
generically suffer from anomalies. We present two different ways that allow 
one to understand these statements.
\end{abstract}

\end{titlepage}

\section{Introduction}

It is well known that discrete symmetries may be anomalous \cite{Krauss:1988zc}.
If this is the case, this can have important consequences for phenomenology. It
implies that the symmetry is violated (at least) at the non--perturbative level.
Originally, anomaly constraints for Abelian finite groups have been derived by
considering \U1 symmetries that get spontaneously broken to
\Z{N}~\cite{Ibanez:1991hv,Ibanez:1991pr,Banks:1991xj}. An arguably more direct
derivation is based on the path integral approach
\cite{Fujikawa:1979ay,Fujikawa:1980eg}, which can also be applied to discrete
symmetries~\mbox{\cite{Araki:2006mw,Araki:2008ek}}. In this approach, a given 
symmetry operation is said to be anomalous if it implies a non--trivial 
transformation of the path integral measure. From this it is straightforward 
to see that there are no cubic anomalies for global symmetries. We can hence 
limit our discussion to anomalies of the type $D-G-G$, where $D$ denotes the 
discrete symmetry and $G$ the continuous gauge group of the setting, 
respectively.

An alternative approach to ensure anomaly freedom is to start  with an
anomaly--free continuous symmetry and breaking it to a discrete
subgroup~\cite{Frampton:1994rk,Luhn:2008sa}. One then obtains embedding
constraints, which guarantee anomaly--freedom but generally are more 
restrictive than the true anomaly constrains. In this work, we use the path
integral approach to discuss anomalies of  discrete symmetries and focus on the
true anomaly constraints.

As noted already in  \cite{Chen:2013dpa}, the $D-G-G$ anomaly
coefficient vanishes if $D$ is a  so--called perfect group because then the
generators of $D$ are traceless, in close analogy to the Lie group case. In this
study, we present a more thorough discussion of the argument. We then present an
alternative argument, based on the observation that the path integral measure
transforms in a one--dimensional representation of the discrete group $D$.
Besides offering an alternative but completely equivalent proof that perfect
groups are anomaly--free, this allows us to conclude that all non--perfect
groups generically have anomalies.  Nevertheless, there exist  particular
non--perfect discrete groups $D$ such that for $G = \SO{N}$ or any  of the
exceptional groups the $D-G-G$ anomaly vanishes independently of the  field
content, and we will give a criterion when this is the case.

Examples for perfect and thus anomaly--safe groups are all non--Abelian finite 
simple groups. This includes, for example, the alternating groups $A_n$ for 
$n\geq5$, the projective special linear groups $\mathrm{PSL}(n,k)$ for $n>1$ 
and finite fields $k$ with more than three elements, and also the sporadic
groups. An example for groups which are not  simple yet perfect and
anomaly--safe are the special linear groups  $\mathrm{SL}(n,k)$ with $n>1$ and
$k>3$ \cite{Wilson:2009}. Furthermore, the (semi--)direct product of two 
perfect groups is again a perfect group (as proven in \Appref{app:groups}).

On the other hand, all non--perfect groups generically can suffer from anomalies. 
For example, this includes $A_4$, $\mathrm{T}'$, $\mathrm{T}_7$, $S_n$, $D_n$, 
which have been utilized frequently in model building, and in general all 
groups that have at least one non--trivial one--dimensional representation 
(cf.\ \cite{Ishimori:2010au} for an extensive list of discrete groups).

As we shall also discuss, anomalies of finite groups can always be cancelled by
a discrete version of the Green--Schwarz (GS) mechanism~\cite{Green:1984sg}.
However, in this case the symmetry is not exact, i.e.\ there exist certain 
terms that violate it.

\section{Anomalies of discrete groups}

Let us start by discussing a quantum field theory with a finite discrete
symmetry $D$. For definiteness, we consider the case that there is also a
non--Abelian gauge symmetry $G$, noting that our arguments also hold for
Abelian gauge factors and gravity.

Furthermore, we assume that there is a set of Dirac fermions $\Psi$ charged
under $D$ and transforming in a representation $\rep{R}$ under $G$.
Given an element $\mathsf{u}\in D$ let $U_{\rep{r}}(\mathsf{u})$ be the unitary 
representation matrix of $\mathsf{u}$ in the unitary representation $\rep{r}$.
For finite groups, there always exists an integer $M_{\mathsf{u}}$ such that 
$\mathsf{u}^{M_{\mathsf{u}}}=\mathsf{e}$. This allows us to write 
\begin{equation}\label{eq:lambda}
 U_{\rep{r}}(\mathsf{u})
 ~=~
 \mathrm{e}^{2\pi\,\I\,\lambda_{\rep{r}}(\mathsf{u})\,/\,M_\mathsf{u}}\;,
\end{equation}
with a matrix $\lambda_{\rep{r}}(\mathsf{u})$ that has integer eigenvalues. Let
us now investigate a discrete chiral transformation under which the
left--handed fermion fields $\Psi_\mathrm{L}:=\PL\Psi$ transform as
\begin{equation}
 \Psi_\mathrm{L}~\to~U_{\rep{r}}(\mathsf{u})\,\Psi_\mathrm{L}
 ~=~
 \mathrm{e}^{2\pi\,\I\,\lambda_{\rep{r}}(\mathsf{u})\,/\,M_\mathsf{u}}\,\Psi_\mathrm{L}\;,
\end{equation}
where $\PL$ is the left--chiral projector and $\rep{r}$ is the representation of
$\Psi_\mathrm{L}$ under $D$. 

The transformation of fermion fields induces, in general, a transformation of 
the path integral measure 
\begin{equation}
 \mathcal{D}\Psi\,\mathcal{D}\overline{\Psi} ~\to~
 J_{\Psi}^{-2}\,\mathcal{D}\Psi\,\mathcal{D}\overline{\Psi}
\end{equation} 
with possibly non--trivial Jacobian $J_\Psi$. For the set of fields $\Psi$ the
Jacobian under the transformation $\mathsf{u}$ is given by
\begin{equation}\label{eq:Jacobian}
 J_\Psi^{-2}~=~\exp\left\lbrace \I\,\frac{2\pi}{M_{\mathsf{u}}}\,\tr[\lambda_{\rep{r}}({\mathsf{u}})]\,\cdot\,\ell(\rep{R})\,\cdot\,
 \int\!\D^4x\,\frac{1}{16\pi^2}\,F^{{a},\mu\nu} \widetilde{F}^{a}_{\mu\nu}\right\rbrace\;.
\end{equation}
Here, $F_{\mu\nu}:= F_{\mu\nu}^a\mathsf{T}_a$ denotes the field strength
tensor of the gauge group $G$ with generators $\mathsf{T}_a$, and 
$\widetilde{F}^{\mu\nu}:=\frac{1}{2}\varepsilon^{\mu\nu\rho\sigma}F_{\rho\sigma}$
its dual. Our conventions are such that $F_{\mu\nu}:=\I
\left[D_\mu,D_\nu\right]$ for the covariant derivative $D_\mu:=\partial_\mu-\I
A_\mu$.

The Dynkin index of the corresponding gauge group representation 
$\ell(\rep{R})$ is defined as usual,
\begin{equation}
 \delta_{ab}\,\ell(\rep{R})
 ~:=~
 \tr\left[\mathsf{T}_a(\rep{R})\,\mathsf{T}_b(\rep{R})\right]\;.
\end{equation}
We fix the Dynkin index following the conventions of \cite{Bernard:1977nr}.\footnote{%
This amounts to normalizing the length of the longest root to unity. The Dynkin 
index of the adjoint representation is then the same as the dual Coxeter number 
of the group.} For the simple compact Lie groups, the resulting Dynkin index 
$\ell(\rep{F})$ of the fundamental representation \rep{F}, which is always 
taken to be (one of) the smallest dimensional representation(s), is shown in
\Tabref{tab:DynkinIndices}.

\begin{table}[t]
\begin{center}
\renewcommand{\arraystretch}{1.25}
\begin{tabular}{lcccccccc}
  $G$ \hspace{5cm}& \SU{N} & $\mathrm{Sp}(N)$ & \SO{N} & $\mathrm{G}_2$ & $\mathrm{F}_4$ & $\mathrm{E}_6$ & $\mathrm{E}_7$ & $\mathrm{E}_8$\\
  $\ell(\rep{F})$ & $\nicefrac{1}{2}$ & $\nicefrac{1}{2}$ & $1$ & $1$ & $3$ & $3$ & $6$ & $30$ \\
\end{tabular}
\end{center}
\caption{Dynkin indices of the fundamental representations \rep{F} for the simple compact Lie groups.}
\label{tab:DynkinIndices}
\end{table}

We define
\begin{equation}
 p~:=~\int\!\D^4x\,\frac{1}{32\pi^2}\,F^{a,\mu\nu} \widetilde{F}^a_{\mu\nu}\;
\end{equation}
which in our convention is an integer \cite{Belavin:1975fg,Bernard:1977nr} in order to simplify \Eqref{eq:Jacobian} to
\begin{equation}\label{eq:Jacobian2}
  J_\Psi^{-2}~=~\exp\left\lbrace \I\,\frac{2\pi}{M_{\mathsf{u}}}\,p\,\cdot\,\tr[\lambda_{\rep{r}}({\mathsf{u}})]\,\cdot\,2\,\ell(\rep{R})\right\rbrace\;.
\end{equation}
When performing the gauge--field path integral, i.e.\ integrating over all
gauge--field configurations, $p$ assumes all integer values. Therefore, we have 
to discuss the anomaly independently of the exact value of $p$ and can take 
advantage only of the fact that it is integer.

In case there are multiple fermions in the theory, their contribution to the 
path integral measure is the product of their respective Jacobians. This 
amounts to summing up the individual contributions in the exponential. Thus, 
the overall effect on the path integral measure due to a transformation 
$\mathsf{u}$, which generates a cyclic group $\Z{M_{\mathsf{u}}}$, can be 
summarized by defining the anomaly coefficient
\begin{equation}\label{eq:A_G-G-ZNR}
 A_{G-G-\Z{M_{\mathsf{u}}}}
 ~:=~
 \sum_f \tr[\lambda_{\rep{r}^{(f)}}({\mathsf{u}})]\,\cdot\,2\,\ell(\rep{R}^{(f)})\;.
\end{equation}
Here, the sum runs over all chiral fermions $f$ transforming in representations
$\rep{r}^{(f)}$ under $D$ and in representations $\boldsymbol{R}^{(f)}$ under
$G$. Note that by the inversion of \Eqref{eq:lambda} one obtains
\begin{equation}\label{eq:delta}
 \tr[\lambda_{\rep{r}}(\mathsf{u})]
 ~=~
 \frac{M_\mathsf{u}}{2\pi\,\I}\,\ln\,\det\,U_{\rep{r}}(\mathsf{u})\;.
\end{equation}
Since the trace of $\lambda_{\rep{r}}(\mathsf{u})$ is only fixed modulo
$M_{\mathsf{u}}$ due to the multi--valued logarithm and because
$2\,\ell(\rep{R})$ is integer, $A_{G-G-\Z{M_{\mathsf{u}}}}$ is only defined
modulo $M_\mathsf{u}$. 

In general it is possible that 
\begin{equation}\label{eq:anomaly_condition}
A_{G-G-\Z{M_{\mathsf{u}}}}~\neq~0\mod M_{\mathsf{u}}\;,
\end{equation}
implying that the overall Jacobian $J$ is different from one and the group
generated by $\mathsf{u}$ is anomalous.

\subsubsection*{Perfect groups are anomaly--safe}

Let us now consider the particular case that the group $D$ is a perfect group. A
perfect group, by definition, equals its commutator subgroup, see also
\Appref{app:groups}. As such, all generating elements $\mathsf{d}\in D$ of the
group (but, in general, not all elements) can be written as the
(group--theoretical) commutator 
\begin{equation}
 \mathsf{d}~=~\left[\mathsf{v},\mathsf{w}\right]~:=~\left(\mathsf{v}\,\mathsf{w}\,\mathsf{v}^{-1}\,\mathsf{w}^{-1}\right)\;,
\end{equation}
of some group elements $\mathsf{v,w}\in D$. This implies that any group element 
$\mathsf{u}\in D$ can be written as a product of commutators
\begin{equation}
 \mathsf{u}~=~\prod_i\,\left[\mathsf{v}_i,\mathsf{w}_i\right]\;,
\end{equation}
where $\mathsf{v}_i,\mathsf{w}_i \in D$. Irrespective of the particular
representation, any representation matrix can thus be written as
\begin{equation}
  U_{\rep{r}}(\mathsf{u})~=~\prod_i\,\left(U_{\rep{r}}(\mathsf{v}_i)\,U_{\rep{r}}(\mathsf{w}_i)\,{U_{\rep{r}}(\mathsf{v}_i)}^{-1}\,{U_{\rep{r}}(\mathsf{w}_i)}^{-1}\right)\;.
\end{equation}
This shows that $\det U_{\rep{r}}(\mathsf{u})=1$, implying that the generator
matrix $\lambda_{\rep{r}}(\mathsf{u})$ in \Eqref{eq:delta} is traceless for 
all representations \rep{r} of the perfect group $D$. Therefore, no element of
a perfect group can give rise to a non--trivial anomaly coefficient. From this
we conclude that for perfect groups all anomalies vanish \cite{Chen:2013dpa}. 

\subsubsection*{The Jacobian as a one--dimensional representation of
\boldmath $D$}

Let us now discuss a possibly more intuitive way to arrive at the same
conclusion. Using \Eqref{eq:delta}, the Jacobian \eqref{eq:Jacobian2} can be
written as
\begin{equation}\label{eq:Jacobian3}
 J_\Psi^{-2} ~=~ \det{\left(U_{\rep{r}}(\mathsf{u})\right)}^{2\,\ell(\rep{R})\,\cdot\,p}\;,
\end{equation}
which might be more useful than \eqref{eq:Jacobian2} for finite groups since it
does not refer to the generators but directly uses the representation matrices
to express the anomaly. Thus, a transformation $\mathsf{u}$ is anomaly--free if
and only if
\begin{equation}\label{eq:anomaly_free_det}
  \prod_f\, \det{\left(U_{\rep{r}^{(f)}}(\mathsf{u})\right)}^{2\,\ell(\rep{R}^{(f)})}~=~1\;.
\end{equation}

Note that the determinant of any representation is a well--defined 
one--dimensional representation because
\begin{align}
  \det{\left(U_{\rep{r}}(\mathsf{u\,v})\right)} ~=~ \det{\left(U_{\rep{r}}(\mathsf{u})\,U_{\rep{r}}(\mathsf{v})\right)} ~=~ \det{\left(U_{\rep{r}}(\mathsf{u})\right)}\,\det{\left(U_{\rep{r}}(\mathsf{v})\right)}\,,
\end{align}
and, furthermore, any integer power of a one--dimensional representation is
again a one--dimensional representation. 

Since the exponent in \eqref{eq:Jacobian3} is integer, we conclude that 
$J_\Psi^{-2}$ transforms in a one--dimensional representation of $D$. In case 
there are multiple fermions, the transformation of the total path integral 
measure $J^{-2}$ is obtained as the direct product of the single 
one--dimensional representations of the individual Jacobians, which is again a 
well--defined one--dimensional representation of $D$. 

The statement that perfect groups are free of anomalies can now be understood 
in a different but completely equivalent way. One can show (for a proof see 
\Appref{app:groups}) that the following statements are equivalent:
\begin{itemize}
 \item[(i)] a finite group $D$ is perfect.
 \item[(ii)] $D$ has exactly a single one--dimensional representation, namely the trivial one.
\end{itemize} 
Furthermore, by the arguments laid out above, the path integral  measure always
transforms in a one--dimensional representation. Thus, for settings based on
perfect groups, the path integral  measure can only transform in the trivial
representation, i.e.\ it does not transform at all and perfect groups are 
anomaly--safe.

Let us remark that the absence of non--trivial one--dimensional representations 
for perfect groups implies that they cannot be used as non--Abelian discrete 
$R$ symmetries with $\mathcal{N}=1$ supersymmetry~\cite{Chen:2013dpa}. 
Moreover, model building (e.g.\ for flavor physics) with perfect groups is 
generally more restrictive because potentials with only multi--dimensional 
representations tend to be more constrained.

\subsubsection*{Non--perfect groups and anomalies}

Consider now the case of a discrete group $D$ which is not perfect. It follows
from the above equivalence that non--perfect groups  always have at least one
non--trivial one--dimensional representation. Consequently, theories based on
non--perfect groups can be anomalous depending on the specific field content,
i.e.\ non--perfect groups are, in general, not safe from anomalies.

However, for some non--perfect discrete groups combined with \SO{N} or
exceptional gauge groups, anomaly freedom is automatic, independently of the
field content. That is, there are some discrete groups $D$ for which the mixed
$D-G-G$ anomalies always cancel if $G$ is an \SO{N} or exceptional group but not
if $G=\SU{N}$. Let us discuss a general criterion when this is the case. For
\SO{N} or exceptional gauge groups, the Dynkin index $\ell(\rep{F})$ is not
$\nicefrac{1}{2}$ but some integer, cf.\ \Tabref{tab:DynkinIndices}. Therefore,
a generic setting based on such a gauge group and a non--perfect finite group 
is anomaly--free as \eqref{eq:anomaly_free_det} is satisfied, provided that the 
finite group exclusively has non--trivial one--dimensional representations 
$\rep[_\mathrm{nt}]{1}$ which obey
\begin{equation}\label{eq:rootOfUnity}
 \left(\rep[_\mathrm{nt}]{1}\right)^{2\,\ell(\rep{F})}~=~\rep[_0]{1}\;,
\end{equation}
where $\rep[_0]{1}$ is the trivial one--dimensional representation.
An example is provided by the symmetric groups $S_n$, which have only one
non--trivial one--dimensional representation with some elements represented as
$-1$. Hence, for symmetric groups the $S_n-G-G$ anomaly vanishes for 
$G$ not being \SU{N} or $\mathrm{Sp}(N)$ independently of the field content.

More generally, \Eqref{eq:rootOfUnity} is certainly fulfilled for any
combination of discrete group $D$ and gauge group $G$ for which
\begin{align}
  \frac{2\,\ell(\rep{F})}{|D/[D,\,D]|} \in \Z{}\;.
\end{align}
That is, it is fulfilled if the order of the Abelianization of $D$ divides
twice the smallest Dynkin index of $G$. This criterion can be further
refined. To see this, note that, using the fundamental theorem of finite Abelian
groups (cf.\ e.g.\ \cite{Gallian:2012}), the Abelianization $D/[D,\,D]$ can
always be written in standard form as a direct product of \Z{p_i^{\nu_i}}
factors where each order $p_i^{\nu_i}$ is some power $\nu_i$ of
a prime number $p_i$.  Thus, the maximal order of elements of $D/[D,\,D]$ is the
least common multiple of the $p_i^{\nu_i}$ (cf.\ e.g.\ \cite{Gallian:2012}).
Hence, the group is anomaly--free with respect to $G$ independently of the field
content if and only if the least common multiple of the $p_i^{\nu_i}$ divides
$2\,\ell(\rep{F})$.

However, we see that in general non--perfect groups are not safe from anomalies.
As usual, anomaly freedom amounts to imposing constraints on the spectrum and
the continuous gauge symmetry $G$.

\subsubsection*{Further comments}

Let us explain the relevance of our statements for finite simple groups. It is
well known that non--Abelian finite simple groups are perfect, cf.\ e.g.\
\mbox{\cite[p.\ 27]{Ramond:2010zz}} and \Appref{app:groups}.  As such,
non--Abelian finite simple groups are always safe from anomalies. Abelian finite
simple groups, on the contrary, are non--perfect and therefore generically
suffer from anomalies.

Finally, let us also comment on infinite (i.e.\ non--compact) discrete groups. 
By definition, $|D/[D,D]|=1$ also holds for infinite perfect groups. Thus, we 
expect that also infinite perfect groups are anomaly--safe.  Non--perfect 
infinite groups, on the other hand, have at least one non--trivial 
one--dimensional representation such that settings involving such groups may be 
anomalous in general.  An example for a non--perfect infinite group is 
$\text{SL}(2,\Z{})$.  This group appears as $T$--duality symmetry in 
superstring theories. It is known that it may exhibit anomalies 
\cite{Derendinger:1991hq, Ibanez:1992hc}, which actually allow one to draw
interesting conclusions on the properties of the underlying model.

\section{Green--Schwarz cancellation of discrete anomalies}

In the remainder of this study we wish to discuss settings in which  the anomaly
coefficient \eqref{eq:A_G-G-ZNR} is non--vanishing. Yet the anomaly, i.e.\ the
transformation of the path integral measure, may be compensated by  a
corresponding transformation of an `axion'. This is the Green--Schwarz (GS)
mechanism \cite{Green:1984sg} for discrete symmetries
\cite{Lee:2011dya,Chen:2012jg}. For it to work, the `axion' field $a$ needs to
couple to the corresponding field strength via
\begin{equation}\label{eq:Laxion1}
 \mathscr{L}_{\mathrm{axion}}
 ~\supset~
 - \frac{a}{f_a}\,\left(F_{\mu\nu}\, \widetilde{F}^{\mu\nu}\right)
\end{equation}
with $f_a$ denoting its decay constant. Further, the axion $a$ has to
transform with a shift $a\to a+\Delta_{\mathsf{u}}$ under the anomalous
transformation $\mathsf{u}$. Here, $\Delta_{\mathsf{u}}$ needs to be such that it
precisely cancels the factor in front of $F_{\mu\nu}\,
\widetilde{F}^{\mu\nu}$  in \eqref{eq:Jacobian}. Whereas one can always define such a shift for a single Abelian symmetry, 
the shifts for different Abelian subgroups of a non--Abelian group have to be mutually consistent to cancel the anomaly of the whole group \cite{Chen:2013dpa}.
The fact that the path integral measure $J^{-2}$ transforms in a well--defined one--dimensional representation of
$D$ nicely explains why such a cancellation is always possible.

One may think of the axion $a$ as the complex phase of a field,
$\Phi=r\,\mathrm{e}^{\I\,a}$, which transforms in the complex conjugate
representation of $J^{-2}$. Therefore, there is the possibility of having
allowed operators of the form $\mathrm{e}^{\I\,\beta\,a}\,\mathcal{O}$ with 
some constant $\beta$. Here $\mathcal{O}$ denotes an operator that transforms 
under $D$ with a phase, i.e.\ $\mathcal{O}$ is in a non--trivial 
one--dimensional representation. Without the axion--dependent prefactor, 
$\mathcal{O}$ is hence prohibited by the symmetry. Upon the axion acquiring its 
VEV, the terms of the form $\mathrm{e}^{\I\,\beta\,a}\,\mathcal{O}$ appear to 
violate the discrete symmetry $D$ (similar to the case of a pseudo--anomalous 
\U1, see e.g.\ \cite{Binetruy:1996uv}). That is, unlike for the case of 
anomaly--free discrete symmetries, in the case of pseudo--anomalous discrete 
symmetries there will be terms that may considerably alter the phenomenology of 
models.

To conclude, there are just two possibilities for the consideration of 
anomalies in finite groups: either the group is perfect and the anomalies vanish
automatically, or the group is not perfect. In the second  case, there may be
anomalies depending on the field content; yet one can always  consistently use a
Green--Schwarz mechanism to cancel the anomaly. However, as  mentioned earlier,
the symmetry is then broken by certain (e.g.\  non--perturbative) terms. Hence,
statements concerning phenomenological  consequences of models based on such
pseudo--anomalous symmetries need to be  taken with some care. In particular,
one may be concerned whether or not such symmetry breaking  effects are properly
included.

\section{Conclusion}
\label{sec:Conclusions}

We have shown that non--Abelian finite simple groups, and more generally all
perfect groups, are anomaly--free. Our argument is based on the fact that the
generators of perfect groups are traceless. This argument may also be rephrased
as follows. Due to the fact that the path integral measure corresponding to a
$D-G-G$ anomaly transforms in a one--dimensional representation of $D$, groups
$D$ without non--trivial one--dimensional representations, i.e.\ perfect groups,
cannot have anomalies.

Non--perfect groups, on the contrary, always have at least one non--trivial
one--dimensional representation and therefore are not safe from anomalies in
generic settings. Whether a certain model is anomaly free then depends, as
usual, on the field content. However, as we have seen, under certain
circumstances one can make statements independently of the field content.
Specifically, there are combinations of certain
non--perfect groups and \SO{N} or exceptional gauge groups which are
anomaly--free irrespective of the field content. We have given a criterion when
this is the case.

In the case of a non--vanishing $D-G-G$ anomaly, one can infer the 
representation that a Green--Schwarz axion needs to furnish in order to cancel 
the anomaly directly from the non--trivial representation of the measure.
This also shows that Green--Schwarz anomaly cancellation is always possible 
for finite groups.

Note that our argument is somewhat analogous to the case of non--Abelian
continuous groups. In the case of a global Lie group $L$, it is well known
that $L-G-G$ anomalies cancel because all generators of $L$ are traceless,
exactly like in the case of perfect groups.
One may also attribute this to the fact that Lie groups do not have
non--trivial one--dimensional representations. However, gauged Lie groups are
different in that one also has to consider the $L-L-L$ anomalies. The latter are
not proportional to the trace of a single generator and thus may not be
described by a linear one--dimensional group representation. As is well known,
the corresponding cubic anomaly coefficients do not vanish in general. However,
they always vanish for real representations and, in particular, for the
so--called `safe' groups \cite{Georgi:1972bb}.

\subsection*{Acknowledgments}

M.R.\ would like to thank the UC Irvine, where part of this work was done,
for hospitality. M.--C.C. would like to thank TUM, where part of this work was done, 
for hospitality. This work was partially supported by the DFG cluster of
excellence ``Origin and Structure of the Universe'' (www.universe-cluster.de)
by  Deutsche Forschungsgemeinschaft (DFG), the DFG Research Grant ``Flavor and CP
in supersymmetric extensions of the Standard Model'', the DFG Graduiertenkolleg 1054
``Particle Physics at the Energy Frontier of New Phenomena'' and the TUM Graduate
School. The work of M.--C.C. was supported, in part, by the U.S. National 
Science Foundation (NSF) under Grant No. PHY-1316792 and PHY11-25915. 
M.--C.C., M.R.\ and P.K.S.V.\ would like to thank the Aspen Center for 
Physics for hospitality and support. This research was done 
in the context of the ERC Advanced Grant project
``FLAVOUR''~(267104).

\appendix
\section{Some basic facts about finite groups}
\label{app:groups}

In this appendix, we collect some basic facts in connection to perfect and
simple groups. Further details can be found e.g.\ in \cite{Ramond:2010zz}.

The commutator subgroup $[D,D]$ (also called derived subgroup) of a group 
$D$ is the group which is generated by all commutator elements of $D$, that is
\begin{align}
  [D,\,D] ~:=~\big\langle\mathsf{u}~ : ~\mathsf{u} \in D ~\text{and}~ \mathsf{u} ~=~ \mathsf{v}\,\mathsf{w}\,\mathsf{v}^{-1}\,\mathsf{w}^{-1} ~\text{ for some }~ \mathsf{v},\,\mathsf{w} \in D \big\rangle
  \;.\label{eq:CommutatorSubgroup}
\end{align}
The commutator subgroup is a normal subgroup of $D$, see for example
\cite[p.~27]{Ramond:2010zz}. A perfect group is a group which equals
its own commutator subgroup $D\equiv[D,D]$, or equivalently, for which $|D/[D,D]|=1$.

In what follows, we show that a group is perfect if and only if it has exactly 
a single one--dimensional representation, namely the trivial one. For this, 
note that there is a one--to--one correspondence between the representations of 
a quotient group $D/N$, where $N$ is a normal subgroup of $D$, and certain 
representations of the parent group $D$. In fact, each representation \rep{r} 
of $D$ for which all elements $\mathsf{n} \in N$ are represented by the 
identity, i.e.\
\begin{equation}
\label{eq:repofn}
  U_{\rep{r}}(\mathsf{n}) ~=~ \mathbbm{1} \qquad \forall\, \mathsf{n} \in N\;,
\end{equation}
is also a representation of the quotient group $D/N$. The converse is also 
true: each representation of the quotient group $D/N$ corresponds to a
representation \rep{r} of $D$ with $U_{\rep{r}}(\mathsf{n}) = \mathbbm{1}$ for
$\mathsf{n} \in N$ (cf.\ e.g.\ \cite[p.~41]{Vinberg:2010}).

A particular Abelian quotient group is $D/[D,\,D]$, the so--called 
Abelianization of $D$. Now, consider a one--dimensional representation 
$\rep[_x]{1}$ of $D$. Then, $U_{\rep[_x]{1}}([D,\,D]) = 1$, since complex 
numbers commute, and \Eqref{eq:repofn} is satisfied. Hence, using the 
one--to--one correspondence discussed above,  the one--dimensional
representation $\rep[_x]{1}$ of $D$ is also a  one--dimensional representation
of the Abelian quotient group $D/[D,\,D]$.  Furthermore, note that for an
Abelian finite group the number of  one--dimensional  representations equals the
order of the group.

Consequently, $D$ and $D/[D,\,D]$ have exactly the same number of 
one--dimensional representations, which in turn is equal to 
$\left|D/[D,\,D]\right|$,
\begin{align} 
    &\text{\# of one--dimensional representations of $D$}\notag\\ 
~=~ &\text{\# of one--dimensional representations of $D/[D,\,D]$} \\
~=~ &\left|D/[D,\,D]\right|\;.\notag
\end{align}
This shows that a discrete group $D$ is perfect, i.e.~$[D,\,D]=D$, 
if and only if it has exactly a single one--dimensional representation, namely 
the trivial one.

A simple group is a group whose only normal subgroups are the group itself and
the trivial subgroup. Since the commutator subgroup $[D,\,D]$ is a normal
subgroup of $D$, there are just two possibilities for the commutator subgroup of
a simple group $D$: either it equals the group, $[D,\,D]=D$, or it is the
trivial group, $[D,\,D]=\{\mathsf{e}\}$. The first case corresponds to
non--Abelian finite simple groups, which are thereby shown to be perfect, and
the second case corresponds to Abelian finite simple groups, which are thereby
not perfect.

A (semi--)direct product of perfect groups is again a perfect group. To show
this, take $D=N\SemiDirect S$ with two perfect groups $N$ and $S$.  Then
every element $\mathsf{d}\in D$ can uniquely be written as
$\mathsf{d}=\mathsf{n}\cdot \mathsf{s}$ for some $\mathsf{n}\in N$ and
$\mathsf{s}\in S$.
Since $N$ and $S$ are perfect groups, each of their elements can be written as 
a product of commutator elements. Therefore, also every element in $D$ can be 
written as a product of commutator elements. Hence, $D$ equals its commutator 
subgroup, and we conclude that $D$ is perfect.

\bibliography{Orbifold}
\addcontentsline{toc}{section}{Bibliography}
\bibliographystyle{NewArXiv} 
\end{document}